\documentclass[twocolumn,superscriptaddress,nobibnotes,nofootinbib,floatfix,aps,prb,citeautoscript,reprint]{revtex4-1}

\usepackage{graphicx,latexsym}
\usepackage{epsfig}
\usepackage{subfigure}
\usepackage{color}
\usepackage{latexsym}
\usepackage{amsmath,bm}
\usepackage{amssymb,amsfonts,mathrsfs,sidecap}
\usepackage{rotating}
\usepackage{mciteplus}
\usepackage{natbib}

\newcommand{\I}{{\rm i}}
\newcommand{\bq}{{\bm q}}
\newcommand{\bG}{{\bm G}}
\newcommand{\bz}{{\bm 0}}
\newcommand{\br}{{\bm r}}
\newcommand{\bk}{{\bm k}}

\newcommand{\etal}{\textit{et al.~}} 

\newcommand{\ilm}{
Institut Lumi\`ere Mati\`ere, UMR5306 Universit\'e Lyon 1--CNRS, 
Universit\'e de Lyon 69622 Villeurbanne cedex, France}

\newcommand{\etsf}{European Theoretical Spectroscopy Facility}

\begin{document}

\title{Strong renormalization of the electronic band gap due to
  lattice polarization in the $GW$ formalism}

\author{Silvana Botti} 
\affiliation{\ilm} 
\affiliation{\etsf}
\author{Miguel A.\,L. Marques} \affiliation{\ilm} 
\affiliation{\etsf}
\date{\today}

\begin{abstract}
The self-consistent $GW$ band gaps are known to be significantly
overestimated. We show that this overestimation is, to a large extent,
due to the neglect of the contribution of the lattice polarization to
the screening of the electron-electron interaction. To solve this
problem, we derive within the $GW$ formalism a generalized
plasmon-pole model that accounts for lattice polarization. The
resulting $GW$ self-energy is used to calculate the band structures of
a set of binary semiconductors and insulators. The lattice
contribution always decreases the band gap. The shrinkage increases
with the size of the longitudinal-transverse optical splitting and it
can represent more than 15\% of the band gap in highly polar
compounds, reducing the band-gap percentage error by a factor of three.
\end{abstract}
\maketitle

For the past decades we have witnessed a steady increase of the
accuracy of first principles electronic band structure
calculations. At the lowest level we have density functional theory
(DFT), normally with a local-density (LDA) or a generalized-gradient
approximation (GGA) to the exchange-correlation
functional. Unfortunately, the Kohn-Sham calculations strongly
underestimates quasi-particle band gaps, often by more than a factor
of two. At the next step, we have the $GW$ approximation of many-body
perturbation theory~\cite{Hedin1965,Hybertsen1986,Onida2002}. For many
years, the standard practice was to start from a DFT calculation, and
to evaluate perturbatively the $GW$ energy corrections to the
Kohn-Sham band structure. This procedure, which we will refer to as
$GW$@LDA, is justified only when the departure wave functions and band
structure are already close to the quasi-particle ones. It yields a
very good agreement with experimental photoemission data, especially
for standard $sp$ materials.  However, it is well known that $GW$@LDA
fails for many crystals that have $d$ electrons participating in the
states close to the band
gap~\cite{Bruneval2006,Gatti2007,Vidal2010,Vidal2010b}. To solve this
problem one can perform restricted self-consistent (sc) $GW$, using
for example the quasi-particle (QP)
sc$GW$~\cite{Faleev2004,Schilf2006} method or perturbative $GW$ on top
of sc Coulomb hole plus screened
exchange~\cite{Bruneval2006,Bruneval2006-b,Gatti2007,Vidal2010,Vidal2010b}.
The sc$GW$ techniques have the advantage of being independent of the
starting point and of giving accurate results for systems that are not
treatable using standard approximations, at the price of a larger
computational cost.  We should emphasize that these techniques allow
for a precise treatment of materials that were considered out-of-reach
for {\it ab initio} calculations only 10 years ago.

Unfortunately, self-consistent $GW$ calculations give too large band
gaps, especially for polar materials, the simplest example being
probably LiF~\cite{Vidal2010}.  The overestimation of the band gap
within QPsc$GW$ was also discussed in literature for
III-V compounds\cite{Svane2010}, Zn-IV-N$_2$ \cite{Punya2011}, and
transition metal oxides~\cite{TiO2,Schilf2006,Vidal2010}. The
causes for these errors were attributed to the lack of vertex
corrections~\cite{Punya2011} or to the underscreening by the
random-phase approximation~\cite{Svane2010} (RPA), and in both cases a
simple empirical correction was adopted to improve the theoretical
calculations (such as scaling the final self-energy
$\Sigma_{\textrm{xc}}$ by about 80\%, obtaining the so-called $0.8
\Sigma$ approximation~\cite{Punya2011}).

The study of electronic correlation and its effects on the
band-structure of solids has a very long tradition. A much less
studied subject is, instead, the influence of phonons on the
quasi-particle spectrum. Only now, in fact, the high level of accuracy
achieved by electronic structure calculations makes the phonon
contributions larger than the theoretical precision.

There exist two contributions to the band gap coming from
phonons: i) The electron-phonon coupling, that can be handled
according to the Allen-Heine-Cardona theory~\cite{Allen1976} using
second-order perturbation theory within the harmonic and adiabatic
approximations.  One obtains two terms of the same order to be
included in the $GW$ self-energy, known as the Fan and Debye-Waller
terms. ii) The phonon contribution to the frequency-dependent dielectric
function, that can be a sizable component of the total
electron screening in polar compounds.

The electron-phonon coupling terms are responsible for the strong
reduction of the band gap of diamond. In fact, in that case the
zero-point contribution to the band gap was proved by Giustino {\it et
  al.}~\cite{Giustino2010} to be 0.6\,eV. They observed that the
example of diamond is an extreme case, while for most semiconductors
and insulators the zero-point renormalization is as small as
10--50\,meV~\cite{Giustino2010} and can be safely neglected. The Fan
and Debey-Waller terms have also been used to determine the
temperature dependence of the band gap of
semiconductors~\cite{Liu2012}.

In this work we want to focus on the second, usually overlooked,
contribution of phononic nature: the screening due to the coupling of
the electro-magnetic field and the oscillating
dipoles~\cite{Bechstedt2005,Trani2010}. Of course, this contribution
exists in polar materials only (and thus it is nonexistent in
diamond). We should emphasize that by ``polar'' we mean any material
that has non-negligible Born effective charges, leading to a
measurable longitudinal-optical (LO) and transverse-optical (TO)
phonon splitting. Most semiconducting and insulating materials known
actually fit into this category. Indeed, polar materials are
characterized by LO and TO infrared-active phonons, whose excitation
induces macroscopic electric fields in the
crystal~\cite{Grosso2000}. These fields contribute, together with the
electronic screening, to the total screening of the Coulomb potential,
and their strength is proportional to the bond ionicity.

Already in 2005 Bechstedt~\etal\cite{Bechstedt2005} observed that the
lattice polarization affects significantly the dressing of the
quasi-particles in ionic materials, inducing large corrections to the
band gap, that they calculated using a model electronic dielectric
function and a static approximation for the lattice contribution to
the screening. Unfortunately, it was already observed in
Ref.~\onlinecite{Bechstedt2005} that that model always overestimates
the band-gap correction. In this Letter, we take a step further and develop a
fully {\it ab initio} framework to include the lattice contributions
to the screening within the $GW$ scheme. The basic ingredients are,
besides the standard RPA screening of $GW$, the LO and TO frequencies
of the infrared active phonon modes. We then calculate the electronic
band structure of a set of polar materials applying this new scheme,
and compare the results with experimental data and with many-body
results including only electronic correlations.

The lattice polarization plays an important role in the determination
of the total screening of polar compounds, as it is evident from the
difference between $\epsilon_\infty$, the electronic
dielectric function at zero frequency, and
$\epsilon_\textrm{s}$, the total static dielectric
function including contributions from both electrons and infrared
active phonons. The ratio between these quantities is related to
the phonon frequencies at the center of the Brillouin zone,
$\omega_{\textrm{LO}}$ and $\omega_{\textrm{TO}}$, through the
Lyddane-Sachs-Teller relation:
\begin{equation}
\frac{\omega_\textrm{LO}^2}{\omega_\textrm{TO}^2} =
\frac{\epsilon_\textrm{s}}{\epsilon_\infty} \,.
\label{eq:standard-LST}
\end{equation}

The standard treatment of optical phonons in crystals using the
dynamical matrix implicitly assumes that the interatomic interactions
are instantaneous. In the case of polar crystals, the long-range
nature of the Coulomb interaction requires a proper account of
retardation effects. It is easy to understand that the coupling of
phonon waves and electromagnetic waves is effective only for $\bq
\rightarrow \bz$, since the speed of sound is negligible if compared
with the speed of light. By combining a continuous approximation for
the description of the mechanical waves of the optical modes and
Maxwell's equations for electromagnetic waves one obtains the
polariton dispersion curves. For a polar crystal the total dielectric
function, which includes electronic and lattice polarization terms,
then reads (here for simplicity in the scalar form):
\begin{equation}
  \epsilon (\bq \rightarrow \bz, \omega)  =
  \epsilon_\textrm{e}(\bq \rightarrow \bz, \omega) +  \epsilon_\textrm{lat} (\bq \rightarrow \bz, \omega)
  \,,
  \label{eq:epstot}
\end{equation}
where $ \epsilon_\textrm{e}(\bq \rightarrow \bz, \omega)$ is the long
wavelength limit of the electronic contribution to the dielectric
function (calculated, e.g., in the RPA approximation), and
$\epsilon_\textrm{lat}$ is defined as the contribution to the
dielectric function due to lattice polarization. This latter quantity
can be related to the electronic screening by a frequency-dependent
generalization of the Lyddane-Sachs-Teller relation, that reads
\begin{equation}
  \epsilon_\textrm{lat} (\bq \rightarrow \bz, \omega) =
  \epsilon_\textrm{e}(\bq \rightarrow \bz, \omega)
  \frac{\omega_\textrm{LO}^2 - \omega_\textrm{TO}^2}{\omega_\textrm{TO}^2 - (\omega + \I \eta )^2} \,,
\label{eq:epsilon-form1}
\end{equation}
In this last expression, the infinitesimal $\eta \rightarrow 0^+$.

In the following discussion, and to keep the notation of the equations at
a reasonably simple level, we will consider only cubic crystals with
two atoms in the unit cell. However, all the equations can be easily
generalized to arbitrary unit cells with more than two atoms, i.e. more than
one infrared active mode~\cite{Gonze-Lee}. 

%In that case, the dielectric function becomes a tensor and there can be more
%infrared active modes, and therefore \eqref{eq:epsilon-form1} has to
%be generalized as
%\begin{equation}
% \epsilon_\textrm{lat} (\bq \rightarrow \bz, \omega) =
% \epsilon_\textrm{e} (\bq \rightarrow \bz, \omega) \prod_j
% \frac{\omega_{\textrm{LO},j}^2 - \omega_{\textrm{TO},j}^2}{\omega_{\textrm{TO},j}^2 - (\omega +
%   \I \eta )^2} \,.
%\label{eq:epsilon-form2}
%\end{equation}

Equations~\eqref{eq:epstot} and \eqref{eq:epsilon-form1} can be used
directly within the $GW$ framework without any further
approximation. However, the very large difference between typical
phonon (of the order of the meV) and electronic (of the order of the
eV) frequencies makes the use of methods like the contour
deformation~\cite{Lebegue2003} unpractical.  We will therefore resort
to the popular plasmon-pole model~\cite{Hybertsen1986,Godby1989} of
the $GW$ theory in order to derive a practical framework to include
the effects of lattice polarization.

For convenience, we now move to the reciprocal lattice $\bG$ space and
write $\epsilon$ as a $\bG$, $\bG'$ matrix.  The electronic
plasmon-pole model for $\epsilon_{\textrm{e}\:\bG\bG'}^{-1}(\omega,
\bq)$ reads:
\begin{equation}
 \epsilon^{-1}_{\textrm{e}\:\bG\bG'}(\bq, \omega) = \delta_{\bG\bG'} + \frac{\Omega^2_{\bG\bG'}(\bq)}
         {\omega^2 - \left[ \tilde{\omega}_{\bG\bG'} (\bq)
           - \I \eta \right]^2} \,.
\label{eq:PP}
\end{equation}
The two parameters $\Omega^2_{\bG\bG'}(\bq)$ and
$\tilde{\omega}_{\bG\bG'} (\bq)$ can be fixed for example by
calculating $\epsilon_{\textrm{e}\:\bG\bG'}^{-1}(\bq, \omega)$ at
$\omega =0$ and $\omega= \I\omega_{\textrm P}$~\cite{Godby1989}, where
$\omega_{\textrm P}$ is the plasmon frequency. In order to determine
the total screening, in the case of $\bq \rightarrow \bz$, one has to
evaluate $\epsilon^{-1}_{\bG, \bG'}(\bq \rightarrow \bz, \omega)$ by
inverting the sum of the two matrices $\epsilon_\textrm{e}(\bq
\rightarrow \bz, \omega) + \epsilon_\textrm{lat} (\bq \rightarrow \bz,
\omega)$. The way to proceed is to apply the matrix
equality~\cite{Miller1981}
\begin{equation}
(G+H)^{-1} = G^{-1} -  \frac{1}{1+g} G^{-1} H G^{-1} \,, 
\label{eq:inversion}
\end{equation}
where
\begin{equation}
g = \textrm{Tr} \left\{ HG^{-1}\right\} \,,
\end{equation}
and $G$ is an invertible matrix. In our case the matrix $G$ is
$\epsilon_{\textrm{e}\:\bG,\bG'} (\bq \rightarrow \bz, \omega)$
and $H$ is $\epsilon_{\textrm{lat}\:\bG,\bG'}(\bq \rightarrow \bz,
\omega) $. A useful simplification comes from the fact that $H$ has
only one non-zero element, the head of the matrix
$\epsilon_{\textrm{lat}\:\bz \bz}(\bq \rightarrow \bz, \omega)$.  In
the following we will remove the explicit dependence on $\bq$ of the
dielectric matrix elements as it is clear that the lattice
polarization contribution exists only for $\bq \rightarrow \bz$. Using
\eqref{eq:epsilon-form1} in \eqref{eq:inversion} it is easy to show
that
\begin{multline}
  \epsilon^{-1}_{\bG\bG'}(\omega) =
  \epsilon^{-1}_{\textrm{e}\:\bG\bG'} (\omega) -
  \\ \frac{ \epsilon^{-1}_{\textrm{e}\:\bG\bz}(\omega) \epsilon^{-1}_{\textrm{e}\:\bz\bG'}(\omega)}
  {\epsilon^{-1}_{\textrm{e}\:\bz\bz}(\omega)}
  \frac{\omega^2_{\textrm{LO}}-\omega^2_{\textrm{TO}}}{\omega^2_{\textrm{LO}}-(\omega + \I
    \eta)^2} \,.
  \label{eq:eps_tot-1}
\end{multline}

If we now replace $\epsilon^{-1}_{\textrm{e}\:\bG\bG'}(\omega)$ by the
standard plasmon-pole model \eqref{eq:PP}~\cite{Godby1989}, we obtain
our final expression for $\epsilon^{-1}_{\bG\bG'}(\omega)$. The
resulting formulas are quite complex, but can be simplified for the
head and the wings of the inverse dielectric matrix, remembering
that phonon frequencies are much smaller than the frequencies of
electronic excitations. The head of the inverse dielectric matrix
reads:
\begin{equation}
  \epsilon^{-1}_{\bz\bz}(\omega) = 1 +
  \frac{\Omega^2_{\bz\bz}}{\omega^2 - \left(\tilde{\omega}_{\bz\bz} - \I \eta
  \right)^2} - \frac{1}{\epsilon_{\infty}} 
  \frac{\omega^2_{\textrm{LO}}- \omega^2_{\textrm{TO}}}{\omega^2_{\textrm{LO}} - (\omega + \I
    \eta)^2} \label{eq:total_screening_head} \,,
\end{equation}
where we used
\begin{equation}
  \epsilon_{\infty} = \frac{1}{\epsilon^{-1}_{\textrm{e}\:\bz\bz}(\bq \rightarrow \bz, \omega_\textrm{LO}\approx 0)}
\end{equation}

\begin{table}[t]
  \caption{\label{tab:gaps} Calculated band gaps (in eV) for the
    selected crystals using different $GW$ schemes compared with
    experimental (exp) photoemission gaps.  The experimental values of
    $\omega_{\textrm{\textrm{LO}}}$ and
    $\omega_{\textrm{\textrm{TO}}}$ (in mHartree) are also given. The
    experimental references are given in the first column. The labels
    $GW^{\textrm{lat}}$ and  $GW^{\textrm{lat}}$@LDA are abbreviations for sc$GW^{\textrm{lat}}$
    and $GW^{\textrm{lat}}$@QPsc$GW$, respectively. The last two rows represent the mean
    absolute error (MAE) and mean absolute percentage error (MAPE).}
  \centering
  \begin{tabular}{cccccccc}                           
    & $\omega_{\textrm{\textrm{LO}}}$ & $\omega_{\textrm{\textrm{TO}}}$ & $GW$@LDA & $GW^{\textrm{lat}}$ &  QPsc$GW$ & 
   sc$GW^{\textrm{lat}}$  & exp  \\
   \hline
     LiF  \cite{exp-LiF}         &   2.99 &  1.39  &  13.24  & 12.05 & 15.81  &  13.69  & 14.20 \\
     LiCl \cite{exp-LiCl}        &   1.74 &  0.87  &   8.60  & 7.97  & 10.28  &   9.05   &  9.4  \\
     NaCl \cite{exp-NaCl}        &   1.21 &  0.75  &   7.73  & 7.14  &  9.52  &   8.37   & 8.5   \\
     MgO  \cite{exp-MgO}         &   3.29 &  1.82  &   6.97  & 6.38  &  8.94  &   7.71   & 7.7  \\
     AlP  \cite{exp-AlP}         &   2.28 &  2.01  &   2.32  & 2.26  &  2.79  &   2.70   & 2.49  \\
     AlAs \cite{exp-AlAs-GaAs}   &   1.84 &  1.65  &   1.88  & 1.83  &  2.34  &   2.26   & 2.23  \\
     GaAs  \cite{exp-AlAs-GaAs}  &   1.30 &  1.22  &   1.16  & 1.15  &  1.52  &   1.46   & 1.52  \\
   \hline
     MAE  &        &        &   0.60  & 1.04 &  0.74  &   0.18   &      \\
     MAPE &        &        &  11.5\% & 16.3\% &  9.4\% &   3.0\%  &
  \end{tabular}
\end{table}

The matrix elements of the wings have very similar expressions, thanks
to the simplification of $\epsilon^{-1}_{\textrm{e}\:\bz\bz} (\omega)$
appearing at the denominator in \eqref{eq:eps_tot-1}:
\begin{multline}
  \epsilon^{-1}_{\bG \bz} = 
  \frac{\Omega^2_{\bG \bz}}{\omega^2 - \left(\tilde{\omega}_{\bG \bz} - \I \eta
  \right)^2} - 
  \epsilon^{-1}_{\textrm{e}\:\bG\bz}(\omega_{\textrm{LO}}) 
\frac{\omega^2_{\textrm{LO}}- \omega^2_{\textrm{TO}}}{\omega^2_{\textrm{LO}} - (\omega + \I
  \eta)^2} \label{eq:total_screening_wings} \,,
\end{multline}
with an analogous expression for the $\bz \bG$ terms.

The other matrix elements of
$\epsilon^{-1}_{\bG\bG'} (\omega)$ are more complicated: beside the
extra poles at $\omega = \pm \omega_\textrm{LO}$, there appear
terms involving poles at $\omega = \pm \tilde{\omega}_{\bG\bz}$,
$\omega = \pm \tilde{\omega}_{\bz\bG}$, and $\omega = \pm
\tilde{\omega}_{\bz\bz}/\epsilon_\infty$.

We can then insert \eqref{eq:PP} and \eqref{eq:eps_tot-1} in the
expression for the screened Coulomb interaction $W=\epsilon^{-1} v$,
where $v$ is the bare Coulomb interaction.  This allows to evaluate
analytically the convolution integral of the self-energy $\Sigma = \I
GW$
\begin{equation} 
  \Sigma(\bq,\omega) = \frac{\I}{2\pi}
  \int_{-\infty}^{\infty} d\omega' e^{i\eta\omega'} G(\bq,\omega+\omega')W(\bq,\omega')\,,
  \label{eq:mb30}
\end{equation}
remembering that
\begin{equation}
  G (\br, \br', \omega) = 
  \sum_{{\bf k}, j} \frac{\phi_{\bk, j} (\br) \phi^*_{\bk, j}(\br')}{\omega - \varepsilon_{\bk,j} + \I \eta
  \,\textrm{sign}(\varepsilon_{\bk,j} - \mu)}\,,
\label{eq:G0}
\end{equation}
where $\phi_{\bk,j}$ are Kohn-Sham or quasi-particle orbitals,
$\varepsilon_{\bk,j}$ are the corresponding energies, and $\mu$ is the
Fermi level.

The frequency integration \eqref{eq:mb30} is performed using the residue theorem, in strict analogy with the
procedure followed for the electronic screening alone, with the only
difference that more than one pole is present in the generalized
plasmon-pole model that we derived. Once again, the full result is
rather cumbersome, but a relatively simple expression is obtained for
the lattice contribution to the matrix elements of $\Sigma$ when only
the head of the matrix $\epsilon^{-1}_{
  \bG\bG'}(\bq \rightarrow \bz, \omega)$ is retained:
\begin{widetext}
\begin{equation}
  \langle\phi_{\bk',j} | \Sigma_\textrm{lat} |\phi_{\bk', j}\rangle =
  - \frac{4 \pi}{V} \frac{\omega^2_{\textrm{LO}} - \omega^2_{\textrm{TO}}}{\epsilon_{\infty}} \sum_j
  \frac{\theta(\mu - \varepsilon_{\bk',j})}{(\omega - \varepsilon_{\bk',j})^2 - (\omega_\textrm{LO} - \I \eta)^2} 
  \lim_{q \rightarrow \bz} \frac{\tilde{\rho}^*_{\bk', ij}(\bq \rightarrow \bz) \tilde{\rho}_{\bk', ij}(\bq \rightarrow \bz)}{q^2}
  \,,
\label{eq:matrix-elements-head}
\end{equation}
\end{widetext}
with $V$ the volume of the unit-cell and $\tilde{\rho}_{\bk',
  ij}(\bq \rightarrow \bz)  = \lim_{\bq \rightarrow \bz} \int_{-\infty}^{\infty} d\br \phi^*_{{\bk'}, i} 
(\br)\exp^{(- \I \bq \cdot \br)} \phi_{\bk'+ \bq, j} (\br) $. Note that this term only
contributes to the occupied states due to the presence of the
Heaviside function $\theta(\mu - \varepsilon_{\bk',j})$.

\begin{figure}[t]
  \includegraphics[width=0.99\columnwidth,angle=0,clip]{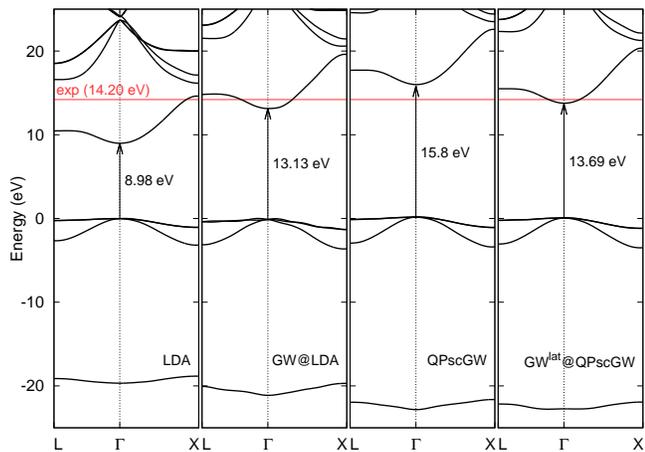}
  \caption{(Color online.) Band structure of LiF along the highest
    symmetry lines of the Brillouin zone.  From left to right:
    Kohn-Sham LDA bands, $GW$@LDA bands, ``standard'' QPsc$GW$ bands,
    and $GW^\textrm{lat}$@QPsc$GW$ bands, including lattice
    polarization in the screening.  The direct band gap value is also
    given. The experimental gap is indicated with a horizontal red
    line.}
  \label{fig:bandsLiF}
\end{figure}

We implemented in {\sc abinit}~\cite{abinit} the complete contribution
of the lattice polarization to $\Sigma$, i.e., the contribution of the
head, wings and body of the dielectric matrix. However, only the poles
at $\omega = \pm \omega_{\textrm{LO}}$ can contribute significantly to
the integral, as the coefficient in front of the other poles is always
extremely small.  For all systems that we studied, the contribution of
the head of the matrix alone is enough to describe the correction to
the band structures within less than 1\,meV, but this may not be the
case for non-cubic systems.

Our calculations start by the determination of the standard Kohn-Sham
ground-state, using the LDA and norm-conserving pseudopotentials. The
LDA Kohn-Sham states are then used to calculate $GW$@LDA band
structures and as a starting point for QPsc$GW$
runs~\cite{Faleev2004}. There are many different versions of $GW$, and
consequently there are many different ways of using
$\Sigma_\textrm{lat}$ in the context of $GW$. We chose as starting
point a converged QPscGW calculation, as this is in our opinion the
most accurate level of theory available in {\sc abinit}. However, one
can not use $\Sigma_\textrm{lat}$ in a self-consistent scheme, as
this would induce contributions of the lattice polarization in
$\epsilon_\textrm{e}$, in disagreement with Eq.~\eqref{eq:epstot}. We
therefore decided to apply the corrections due to the phonons in a
final perturbative step, in what we call the
$GW^\textrm{lat}$@QPsc$GW$ method.

In the following, we present calculations for the band structures of a
set of binary compounds: LiF, LiCl, NaCl, MgO, AlP, AlAs, GaAs.  These
are all polar $sp$ materials, for which perturbative $GW$ gives already good
band gaps, with varied bond ionicities, and band gaps ranging from
1.52\,eV to 14.2\,eV. All calculations were
performed with {\sc abinit}~\cite{abinit}. The energy cutoff was set
to 50\,Ha for LiF, LiCl, NaCl, and MgO, and to 30\,Ha for the
remaining compounds, and the $\bk$-point grids were a
$6\times6\times6$ Monkhorst-Pack. This allowed for a convergence in
the total energy to better than 1\,mHartree. The cutoff for the
dielectric constant was set to 5\,Hartree, and 142 empty states were
used to obtain a convergence in the energy gaps to better than
0.1\,eV.  Finally, and in order to limit the effect of other possible
sources of imprecision, we decided to use for these calculations
experimental lattice constants and experimental LO and TO phonon
frequencies at $\Gamma$ (see Table~\ref{tab:gaps}). Of course the
method can be easily made fully {\it ab initio} by calculating such
quantities within DFT~\cite{abinit}. 

In Table~\ref{tab:gaps} we report the calculated band gaps for the
selected binary compounds using the different $GW$ schemes. As it is
well known, $GW$@LDA underestimates the band gaps of $sp$ compounds by
slightly more than 10\% while QPsc$GW$ overestimates them by about
10\%~\cite{Schilf2006}. Turning on the correction due to lattice
polarization cuts down the error of these techniques by a factor of
three, bringing the mean absolute percentage error to 3\%. These
results clearly prove that the overestimation of the band gap by
QPsc$GW$ is not mainly due to a deficient treatment of electronic
correlation, but to a deficient description of the screening of the
medium. We remark that the inclusion of the lattice contribution to
the screening in a $GW$@LDA calculation deteriorates the agreement
with experiments, leading to a mean absolute percentage error of 16\%.
This shows that the good performance of perturbative $GW$ calculations
for $sp$ polar materials is due to a partial cancellation of errors:
the underestimation of the band gap opening is compensated by the
neglect of the band gap shrinkage due to the lattice polarization.

In Fig.~\ref{fig:bandsLiF} we display the band structures of LiF
obtained with the different approximations under study.  We can
observe that the inclusion of the lattice polarization in the
screening produces a rigid shift downwards of the conduction bands,
with negligible effects on the band widths and band dispersions. While
the same qualitative behavior is observed for all the materials under
study here, we cannot exclude that other effects may be present in
more complex materials.

Finally, we would like to make two remarks: (i)~The coupling between
electrons and phonons in our model is indirect and comes through
Maxwell equations. It is therefore completely unrelated to
electron-phonon coupling. (ii)~To take into account all phonon
contributions to the electronic band structure, both the lattice
contribution to the screening and the Fan and Debye-Waller (and
possibly higher order) terms of the Allen-Heine-Cardona
theory~\cite{Allen1976} should be included. It is true that for the
materials studied here one does not expect a large contribution from
the electron-phonon coupling, but it is not inconceivable to find a
strong ionic compound with strong electron-phonon coupling. Note that
both contributions will tend to decrease the purely electronic gap.

In conclusion, we developed a fully {\it ab initio} $GW$ framework
that includes the effects of the screening owing to the polarization of
the lattice. Within this framework we show that the overestimation of
the band gaps by restricted self-consistent $GW$ techniques is due to
the neglect of this contribution, and we manage to
bring the error in the calculated band gaps to a mere 3\%.  The
lattice contribution decreases the band gap by a factor that increases
with the size of the longitudinal-transverse optical splitting and it
can represent more than 15\% of the band gap in highly polar
compounds.  These results call for a reexamination of many theoretical
calculations of the quasi-particle spectrum for polar materials,
including for many oxides, nitrides, etc. that are important in
several fields of technology.

%==========Acknowledgments 

We would like to thank Julien Vidal and Friedhelm Bechstedt for many
fruitful discussions. We acknowledge financial support
from the French ANR project ANR-12-BS04-0001-02. Computational
resources were provided by GENCI (project x2011096017).

\end{document}